\documentclass{iopart}
\usepackage{graphicx}

\usepackage{latexsym}
\usepackage{amssymb}
\usepackage{color}

\begin{document}

\title[Inertia, Gravity and the Meaning of Mass]
{Inertia, Gravity and the Meaning of Mass}

\author{Fulvio Melia}
\address{Department of Physics, The Applied Math Program, and Department of Astronomy,
The University of Arizona, AZ 85721, USA\\
E-mail: fmelia@email.arizona.edu}

\begin{abstract}
Our concept of mass has evolved considerably over the centuries, most
notably from Newton to Einstein, and then even more vigorously with
the establishment of the standard model and the subsequent discovery
of the Higgs boson. Mass is now invoked in various guises depending on
the circumstance: it is used to represent inertia, or as a coupling
constant in Newton's law of universal gravitation, and even as a repository
of a mysterious form of energy associated with a particle at rest.
But recent developments in cosmology have demonstrated that rest-mass
energy is most likely the gravitational binding energy of a particle
in causal contact with that portion of the Universe within our
gravitational horizon. In this paper, we examine how all these 
variations on the concept of mass are actually interrelated via 
this new development and the recognition that the source of gravity 
in general relativity is ultimately the total energy in the system.
\end{abstract}

\pacno{04.20.-q, 04.20.Ex, 95.36.+x, 98.80.-k, 98.80.Jk}

\section{Introduction}
Already by Newton's time there were potentially two kinds of mass
invoked in burgeoning physical laws. On the one hand, objects
exhibited an acceleration in proportion to the force applied to
them, implying they possessed a conserved `inertial mass,' $m_i$.
And after Newton formulated his law of universal gravitation, it became
apparent that a body also has `gravitational mass,' $m_g$, that
today we would refer to as a gravitational coupling constant. Newton
viewed these two quantities as being conserved, irreducible properties
of matter, and simplified the description further by considering them
to be indistinguishable \cite{Newton:1687}.

Bondi \cite{Bondi:1957} refined these definitions further, 
including also a possible dichotomy between `passive' gravitational
mass--that which responds to a gravitational field--and `active'
gravitational mass--that which creates the gravitational effect. He
also allowed for the possibility of negative values for all these
quantities. He argued, however, that the law of action and reaction
in Newtonian physics implies the equality of active and passive
gravitational masses. This concept has been tested experimentally
many times since then, beginning with Kreuzer \cite{Kreuzer:1968}, who 
inferred an upper bound of $\sim 5\times 10^{-5}$ for the fractional 
difference between the passive and active gravitational masses, to
the latest measurement by Singh et al. \cite{Singh:2023}, who lowered
the limit considerably to $\sim 3.9\times 10^{-14}$.

All we can really say about inertial and gravitational masses 
is that the clues from nature point to a strict proportionality between 
them, since in principle Newton's gravitational constant 
(see Eq.~\ref{eq:Fg} below) can always be adjusted to comply with any 
change in the ratio $m_i/m_g$. This proportionality is also the basis 
for Einstein's Principle of Equivalence, one of the most important 
founding tenets of general relativity \cite{Weinberg:1972}.

A century later, we have a much more nuanced interpretation of mass,
certainly with the establishment of the standard model
\cite{Glashow:1961,Salam:1964,Weinberg:1967} and the subsequent
discovery of the Higgs boson \cite{Englert:1964,Higgs:1964,Oerter:2006}.
We now understand that inertial mass is better described as an emergent
quality rather than an intrinsic property of matter, given that several
fundamental particles, such as electrons, positrons and quarks, owe
their inertia to a coupling with the Higgs field, while other composite
particles, such as neutrons and protons, acquire inertia via the dynamical
back reaction of accelerated quarks, which radiate gluons to conserve
momentum (see ref.~\cite{Melia:2021a} for a recent review). In this context,
$m_i$ ought to be viewed as a `place holder' for something else, a notion
we shall utilize liberally throughout this paper.

The gravitational mass has undergone quite an evolution as well. Following
the broad acceptance of general relativity as the correct description of
space and time, we now view the source of gravity to be the total energy,
$E$, in the system. Coupled to another major discovery in relativity---the
existence of rest-mass energy---it now appears that Newton's gravitational
mass may simply be a more primitive representation of $E/c^2$. None of this
necessarily gives us confidence, though, that $m_i$ and $m_g$ should be 
considered as representing the same thing. 

But more recent work appears to have uncovered the origin of rest-mass
energy \cite{Melia:2021a}, which in the end may tie all of these loose
threads together. As we shall discuss later in this paper, the energy
associated with a particle at rest appears to be its gravitational binding
energy with that portion of the Universe contained within our gravitational
horizon. Thus, in an odd twist of history, $m_i$ and $m_g$ appear to be
linked after all---owing both of their existence to the energy of the particle. 

To unravel this intricate quilt of physical attributes, we shall
examine Newton's law of universal gravitation in the weak-field, static limit
of general relativity, but go beyond its traditional application to particles
`with mass.' We now know that gravity accelerates `massless' particles as 
well, as proven by the measured deflection of light in transit toward Earth
through a cosmic medium with variable gravity. We shall therefore derive
the analog of Newton's law for particles such as photons, which Newton would
never have considered in his gravitational framework for want of any physical 
evidence of their existence. Were he alive today, however, he would no doubt
have devised two versions of Equation~(\ref{eq:Fg}): one for `massive' particles,
the other for `massless.' This will be our principal task in 
\S~\ref{sub.zeroinertia}, with important foundational work in \S~\ref{sub.inertia}.

Finally, in \S~\ref{mgmi}, we shall unify the various concepts of inertial and
gravitational mass via the rest-mass energy associated with them. We shall
conclude with some closing thoughts in \S~\ref{conclusion}.

\section{Gravitational Coupling}\label{gravity}
\subsection{Particles with Established Inertia}\label{sub.inertia}
Newton's law of universal gravitation is an expression of the force experienced by 
a particle with established inertia $m_i$ and gravitational mass $m_g$, due to a 
gravitating mass $M_g$ located at ${\bf r}=0$, 
\begin{equation}
{\bf F}_g=m_i{\bf g}=-{Gm_gM_g\over r^2}\hat{r}\;,\label{eq:Fg}
\end{equation}
which is understood relativistically in the limit of weak, static fields and 
low velocities. Of necessity, the latter condition implies that particles 
coupling to the force must have non-zero inertia, $m_i\not=0$, for the
acceleration {\bf g} would otherwise be infinite. Equation~(\ref{eq:Fg}) may 
also be written in terms of the gravitational potential, 
\begin{equation}
\Phi({\bf r})\equiv -{GM_g\over r}\;,\label{eq:Phi}
\end{equation}
such that
\begin{equation}
m_i{\bf g}=-m_g\vec{\nabla}\Phi\;.\label{eq:gPhi}
\end{equation}

Equations~(\ref{eq:Phi}) and (\ref{eq:gPhi}) actually represent two distinct
phenomena. The first accounts for the potential created by the hypothesized
gravitating mass $M_g$, while the second describes the response of a test particle
$m_g$ to the presence of this potential. Thus, to ensure consistency between
general relativity in the weak-field, static, low-velocity limit and Newton's
law of universal gravitation, two separate physical effects must be considered. 
First, the response of the particle given by Equation~(\ref{eq:gPhi}) is most 
naturally inferred from the geodesic equation, 
\begin{equation}
{d^2x^\mu\over d\lambda^2}+{\Gamma^\mu}_{\alpha\beta}\,{dx^\alpha\over d\lambda}
{dx^\beta\over d\lambda}=0\;,\label{eq:geo}
\end{equation}
describing the trajectory of a {\it free} particle through the spacetime created 
by the gravitating mass $M_g$ (see Eq.~\ref{eq:Schwarzschild} below). The affine 
parameter $\lambda$ is often chosen to
be the proper time $\tau$ in the particle's rest frame when $m_i\not=0$, though
not for massless particles, such as a photon, for which $\tau$ is always zero.
In this expression, ${\Gamma^\mu}_{\alpha\beta}$ are the Christoffel symbols
containing information about the spacetime curvature, most directly calculated 
from the metric coefficients themselves:
\begin{equation}
{\Gamma^\mu}_{\alpha\beta}\equiv {1\over 2}g^{\mu\delta}\left[g_{\beta\delta,\alpha}
+g_{\delta\alpha,\beta}-g_{\alpha\beta,\delta}\right]\;,\label{eq:Chris}
\end{equation}
where the metric is formally written as 
\begin{equation}
ds^2=g_{\alpha\beta}dx^\alpha dx^\beta\;.\label{eq:metric}
\end{equation}

The procedure for reducing Equation~(\ref{eq:geo}) to its Newtonian form is well 
known, so we won't dwell on the details, but mention only the key points, mostly
in preparation for \S~\ref{sub.zeroinertia} below. For a particle with inertia 
moving well below the speed of light, the second term is dominated by the 
$\alpha=\beta=0$ component, and therefore
\begin{equation}
{d^2x^\mu\over d\tau^2}+{\Gamma^\mu}_{00}\,{dx^0\over d\tau}
{dx^0\over d\tau}=0\;.\label{eq:geo1}
\end{equation}
Then, to calculate ${\Gamma^\mu}_{00}$ for a weak gravitational field, we put
\begin{equation}
g_{\alpha\beta}=\eta_{\alpha\beta}+h_{\alpha\beta}\;,\label{eq:gab}
\end{equation}
where $\eta_{\alpha\beta}={\rm diag}(+1,-1,-1,-1)$ and $|h_{\alpha\beta}|\ll 1$. 
The inverse metric tensor is simply 
$g^{\alpha\beta}=\eta^{\alpha\beta}-h^{\alpha\beta}$.
If in addition the field is static,
\begin{equation}
{\Gamma^\mu}_{00}=-{1\over 2}\eta^{\mu\nu}\partial_\nu g_{00}=
-{1\over 2}\eta^{\mu\nu}\partial_\nu h_{00}\,.\label{eq:gm00}
\end{equation}
Substituting Equation~(\ref{eq:gm00}) into (\ref{eq:geo1}), we thus find that
\begin{equation}
{d^2x^\mu\over d\tau^2}={1\over 2}\eta^{\mu\nu}\partial_\nu h_{00}\left({dx^0\over 
d\tau}\right)^2,\label{eq:dxmu}
\end{equation}
so that $d^2x^0/d\tau^2=0$, which means that $dt/d\tau$ is constant. That is,
time progresses forward at a steady rate for a particle in Newtonian gravity, 
consistent with the prevailing view on the nature of time during Newton's era.  

For the spatial coordinates ($j=1,2,3$), we instead find that
\begin{equation}
{d^2x^j\over d\tau^2}=-{c^2\over 2}\left({dt\over d\tau}\right)^2\partial_j h_{00}\label{eq:xj}
\end{equation}
or, using the chain rule of differentiation,
\begin{equation}
{d^2x^j\over dt^2}=-{c^2\over 2}\partial_j h_{00}\;.\label{eq:xjt}
\end{equation}
A comparison of Equations (\ref{eq:gPhi}) and (\ref{eq:xjt}) therefore shows that, in order
for Einstein's theory to correctly describe the motion of a particle with established inertia 
in a classical gravitational field, we must have $m_g\rightarrow m_i$ in the Newtonian 
limit (see \S~\ref{mgmi}) which is, afterall, the basis for the Equivalence Principle 
that gave rise to the general relativistic description of particle trajectories in 
the first place. In addition, $h_{00}\equiv 2\Phi/c^2$, so that
\begin{equation}
g_{00}=1+{2\Phi\over c^2}\;.\label{eq:g00Phi}
\end{equation}

In the next section, we shall learn that the gravitational coupling of particles believed
to have zero inertia is very similar to this result, but differs from it in at least 
one very significant aspect. Before we begin to examine that situation, however, we must
first understand the second phenomenon associated with Newtonian gravity---that giving 
rise to the gravitational potential itself (Eq.~\ref{eq:Phi}). For this, we need to 
begin with the gravitational field equations in general relativity, which may be written
\begin{equation}
G_{\alpha\beta}\equiv R_{\alpha\beta}-{1\over 2}\,g_{\alpha\beta}\,
R=-{8\pi G\over c^4}\,T_{\alpha\beta}\;.\label{eq:Einstein}
\end{equation}
$G_{\alpha\beta}$ is the Einstein tensor, written in terms of the Ricci tensor,
\begin{equation}
R_{\beta\delta}\equiv g^{\alpha\gamma}\,R_{\alpha\beta\gamma\delta}=
R^{\gamma}_{\,\;\;\beta\gamma\delta}\;,\label{eq:Riccitensor}
\end{equation}
the contracted Ricci tensor
\begin{equation}
R\equiv g^{\mu\nu}\,R_{\mu\nu}\label{eq:Ricciscalar}
\end{equation}
(also known as the {\it curvature scalar}) and the stress-energy tensor $T_{\alpha\beta}$.
In Equation~(\ref{eq:Riccitensor}), the quantity
\begin{equation}
{R^\alpha}_{\delta\beta\gamma}\equiv {\Gamma^\alpha}_{\delta\beta\;,\gamma}
-{\Gamma^\alpha}_{\delta\gamma\;,\beta}-{\Gamma^\alpha}_{\epsilon\beta}\,
{\Gamma^\epsilon}_{\delta\gamma}+{\Gamma^\alpha}_{\epsilon\gamma}\,
{\Gamma^\epsilon}_{\delta\beta}\;\;\label{eq:Riemann}
\end{equation}
is known as the Riemann-Christoffel (or {\it curvature}) tensor. 

The appearance of the curvature scalar on the left-hand side of Equation~(\ref{eq:Einstein}) 
is sometimes inconvenient. Contracting this equation with $\alpha$ and $\beta$ reduces it to the form
\begin{equation}
R={8\pi G\over c^4}\,{T^{\gamma}}_\gamma\;.\label{eq:contract}
\end{equation}
Thus, an alternative representation of the field equations is
\begin{equation}
R_{\alpha\beta}=-{8\pi G\over c^4}\left(T_{\alpha\beta}-{1\over 2}\,g_{\alpha\beta}\,
{T^\gamma}_\gamma\right)\;.\label{eq:Einstein1}
\end{equation}
It is beyond the scope of the present paper to describe how and why these equations are 
derived, but this topic is well covered in both the primary and secondary literature 
\cite{Weinberg:1972,Melia:2020}.  We do point out, however, that the coefficient 
multiplying $T_{\alpha\beta}$ on the right-hand side of Equations~(\ref{eq:Einstein}) 
and (\ref{eq:Einstein1}) was chosen in order for Einstein's equations to correctly 
reproduce the Newtonian potential in Equation~(\ref{eq:Phi}) for a gravitating 
inertial source $M_g$ in the weak-field, static, low-velocity limit, which we 
now describe.

For simplicity, we adopt the so-called {\it perfect fluid} approximation, in which the 
stress-energy tensor excludes all possible shear forces associated with the transport 
of momentum components in directions other than those associated with the components 
themselves. The covariant form of this tensor may be written 
\begin{equation}
T_{\alpha\beta}={1\over c^2}\left(\rho+p\right)u_\alpha u_\beta-
p\,g_{\alpha\beta}\;,\label{eq:Tab}
\end{equation}
where $u_\alpha$ is the local value of $dx_\alpha/d\tau$ for a comoving
fluid element in the source, and $p$ and $\rho$ are the pressure and energy density, 
respectively, measured by an observer in a locally inertial frame comoving with the 
fluid at the instant of measurement. 

The trace of this stress-energy tensor is simply
\begin{equation}
T\equiv {T^\gamma}_\gamma=\rho-3p\;,\label{eq:trace}
\end{equation}
providing a clear, unequivocal affirmation that {\sl the source of spacetime curvature 
in general relativity is all forms of energy and momentum.} This aspect of Einstein's
theory cannot be overstated because it represents a clear departure from the Newtonian
framework, in which gravity is due to an intrinsic `mass' associated with the source,
considered by Newton to be synonymous with inertia, and neither having anything to do
with energy. Our continued examination of the meaning of $m_g$, $m_i$ and $M_g$ below 
will be heavily based on this crucial distinction between Einstein's and Newton's
theories, and their behavior in the classical limit. Equation~(\ref{eq:Einstein1}) 
may thus also be written in the more suggestive form
\begin{equation}
R_{\alpha\beta}=-{8\pi G\over c^4}\left(T_{\alpha\beta}-{1\over 2}\,g_{\alpha\beta}\,
[\rho-3p]\right)\;.\label{eq:Einstein2}
\end{equation}

This expression is completely valid for all forms of energy, with or without inertia, 
and we shall use it also in the following section where we consider the gravitational 
coupling of particles believed to have no inertia. Here, however, we use the fact 
that matter (or its more common designation as `dust') has essentially zero pressure, 
so $T=\rho$, and Equation~(\ref{eq:Einstein2}) therefore implies that, to first order 
in the weak-field (Newtonian) limit (see Eq.~\ref{eq:gab}),
\begin{equation}
R_{00}=-{4\pi G\over c^4}\rho\;,\label{eq:Einstein3}
\end{equation}
which relates derivatives of the metric coefficients (specifically $h_{00}$
and $\Phi$) to the energy density, providing a direct link to the Poisson equation
for $\Phi$, from which Equation~(\ref{eq:Phi}) is derived.

From Equation~(\ref{eq:Riccitensor}), we see that
\begin{equation}
R_{00}={R^\gamma}_{0\gamma 0}\;,\label{eq:R00}
\end{equation}
and ${R^0}_{000}=0$, so 
\begin{equation}
R_{00}={R^i}_{0i0}=\partial_i{\Gamma^i}_{00}-\partial_0{\Gamma^i}_{i0}
+{\Gamma^i}_{i\gamma}{\Gamma^\gamma}_{00}-{\Gamma^i}_{0\gamma}{\Gamma^\gamma}_{i0}\;.\;\;\label{eq:R001}
\end{equation}
The second term on the right-hand side is zero for a static field, while the third
and fourth terms are second order in $h_{00}$. This leaves
\begin{equation}
R_{00}=\partial_i{\Gamma^i}_{00}=-{1\over 2}\eta^{ij}\partial_i\partial_j\,h_{00}\label{eq:R002}
\end{equation}
and, combining Equation~(\ref{eq:Einstein3}) with (\ref{eq:R002}), gives
\begin{equation}
\vec{\nabla}^2\Phi = 4\pi G\left({\rho\over c^2}\right)\;,\label{eq:Poisson}
\end{equation}
which is the relativistically derived Poisson's equation for the gravitational 
potential in the presence of an energy density $\rho$ in the weak-field, static limit.
It must be emphasized that this equation excludes any momentum (and therefore the
pressure this would create) of the gravitating source specifically because we
attributed inertia to the medium, allowing it to reside near the origin with
a very low (or even zero) velocity. It is for this reason that $T=\rho-3p$ 
simply reduces to $\rho$ in Equation~(\ref{eq:Einstein2}), and we acknowledge 
the fact that the coefficient $4\pi G$ in Equation~(\ref{eq:Poisson}) was chosen 
to ensure that Einstein's and Newton's theories yield the same potential, $\Phi$, 
when the gravitational mass, $M_g$, is calculated solely from the energy density 
in the system, and nothing else. 

But there is another subtle, yet crucial, feature of this equation that we must
fully understand, particularly when we begin to compare this result
with its counterpart in the following section, addressing gravity in the
presence of energy with what we believe to be zero inertia, i.e., in cases where 
$T=\rho-3p$ is not merely $\rho$. Equation~(\ref{eq:Poisson}) is fully consistent
with Equation~(\ref{eq:Phi}) only if we follow Newton in attributing the gravitational
influence to a hypothesized `gravitational mass,' $M_g$, which necessarily would have
to correspond to a gravitational mass density $\rho_g\equiv\rho/c^2$ in 
Equation~(\ref{eq:Poisson}). 

Those of us accustomed to the language of general relativity would be tempted to
consider this as being self-evident. After all, isn't rest-mass energy simply given by
this relation?  Well yes, but not completely, as we shall soon find out. Later in this 
paper we shall better understand the distinction between gravitational and inertial mass, 
and realize that the interpretation of $M_g$ in Equation~(\ref{eq:Phi}) as the 
`rest-mass'---from the conversion of $\rho$ to $\rho_g\equiv \rho/c^2$ in
Equation~(\ref{eq:Poisson})---is valid only because Newton's law of universal gravitation 
was specifically formulated for particles with non-zero inertia in the low-velocity
limit, where the gravitational and inertial masses are proportional to each 
other---or even equal, with an appropriate choice of units for the gravitational 
constant $G$. The situation is very different for photons, because $m_g$ for them is 
{\bf not} zero. 

\subsection{Gravity with `Zero Inertia'}\label{sub.zeroinertia}
Whether light has inertia and/or gravitational mass, and whether these two are
equal, was something that could not easily be discussed or explored prior to
the advent of relativity theory. But this did not prevent classical physicists
from entertaining the idea that large heavenly bodies, such as the Sun, could in
principle accelerate rays of light and cause them to deviate in discernible ways
from straight-line trajectories. Very famously, Einstein himself used 
his Principle of Equivalence couched in Newtonian theory, essentially 
Equation~(\ref{eq:Fg}) with $m_i$ and $m_g$ cancelled from both sides, to predict 
how much starlight would be bent upon grazing the surface of the Sun on its way 
toward Earth \cite{Einstein:1911}. 

Without the full theory of general relativity to support this calculation, he
partially relied on intuition, arguing that the rate of proper time (as seen by
an observer fixed with respect to the source of gravity) varies with distance
from the center of the gravitating body, thereby creating a speed of light
varying with radius if one ignores the spatial variations. The latter assumption
is key to understanding why he had to correct his prediction once general
relativity was completed five years later. Applying Huygens’s principle to a 
wave front passing through such a region, he could then calculate the degree
of bending based on the time dilation produced by the central mass. This 
prediction amounted to about $0.875$ seconds of arc, which turns out to be 
wrong by a factor $2$. We shall see below why ignoring the spatial variations
results in this not insignificant mistake. One may still predict the correct
deflection angle using Newtonian theory, however, but only by using an 
alternative form of Equation~(\ref{eq:Fg}) appropriate for light 
(see Eq.~\ref{eq:Newtonnull}), which was not available to him at that
time.

This type of speculation had already been carried out by others before
him, even by Newton who, in his treatise on Opticks\cite{Newton:1704} 
published in 1704, asked the question: ``Do not Bodies act upon Light at a distance, 
and by their action bend its Rays, and is not this action strongest at the least 
distance?" A century later, Johann Georg von Soldner had used Newton's Law of
universal gravitation to calculate the deflection of starlight by the Sun treating 
``a light ray as a heavy body'' and predicted a deflection angle of $0.84$ arcseconds, 
virtually identical to Einstein's estimate based on his first attempt \cite{vonSoldner:1821}. 

Einstein redid his calculation five years later once General Relativity was completed, 
taking into account all spacetime curvature effects and corrected his mistake, predicting 
a deflection angle of $1.745$ arcseconds. As is well known by now, Sir Arthur Eddington 
subsequently led an expedition to an island off the coast of Africa, with a second group 
in Brazil, to measure the deflection of starlight grazing the edge of the Sun during the 
total eclipse of May 29, 1919. Their measurement provided a spectacular (if controversial) 
confirmation that General Relativity is the correct theory of gravity \cite{Eddington:1919}. 

Questions have been raised about Eddington's analysis of their data 
because turbulence in Earth's atmosphere causes deflections of starlight comparable
to those predicted by Einstein's theory. One must rely on the assumption that these
are random in nature, so that they can be averaged away with the use of many images,
leaving only the relativistic effect. By the end of their observations, Eddington and
his coworkers had only two reliable images (with about five stars) at one site, and
eight usable plates (with at least seven stars) at the Sobral location. Nineteen
other plates taken with a second telescope had to be abandoned. Given this paucity
of measurements, did Eddington really have the evidence to support Einstein's
theory \cite{Earman:1980}? Some have argued that Eddington's enthusiasm for
general relativity biased his approach. But many re-analyses between 1923 and 1956 
of the plates from those expeditions yielded similar results within ten percent. 
A reanalysis in 1979 using the Zeiss Ascorecord and its data reduction software 
\cite{Harvey:1979} yielded the same deflection as that calculated by
Eddington, though with even smaller errors. The Sobral plates gave similar
results, all consistent with general relativity. A modern assessment of
Eddington's work has thus tended to show no credible evidence of bias
in his conclusions \cite{Kennefick:2009}.

Having said this, we now know, that photons do not have rest mass in the conventional form 
seen in the standard model of particle physics. So why are we justified in using General 
Relativity to calculate the gravitational acceleration of a photon when Einstein's theory is 
based on the equality of $m_i$ and $m_g$ (i.e., the Principle of Equivalence)? This question
will become even more acute later in this paper, when we learn that Newton's approach 
of identifying gravitational mass suggests that $m_g$ strictly cannot be zero for light. 
To place this in context, we should ask ourselves whether we are missing a law of
universal of gravitation, analogous to Equation~(\ref{eq:Fg}), representing the weak-field, 
static limit of General Relativity for photons and other particles that do not have
inertia in the conventional sense. 

Such particles do not follow trajectories described by Equation~(\ref{eq:geo1}).
The magnitude of their velocity is always $c$, so one cannot adopt an asymptotically 
small speed to go along with the assumption of weak, static fields. As was the case in
\S~\ref{sub.inertia}, there are two effects we need to consider: the first arises from
the impact of non-zero momentum on the source of gravity, modifying Newton's definition
of $M_g$, and the second provides the gravitational coupling of a particle we believe
to have zero rest mass to the `force' created by the former.  

The first of these phenomena is quite straightforward to understand. Whereas the
trace of the stress-energy tensor, $T=\rho-3p$, reduces to $\rho$ for `dust,' the 
pressure cannot be ignored when the source is radiation since its momentum makes a 
significant contribution to the overall energy budget. For example, isotropic radiation 
exerts a pressure $p=\rho/3$, so $T=0$. Thus, instead of Equation~(\ref{eq:Einstein3}),
we now get
\begin{equation}
R_{00}=-{8\pi G\over c^4}\rho\;,\label{eq:Einstein4}
\end{equation}
resulting in the modified Poisson's equation
\begin{equation}
\vec{\nabla}^2\Phi = 4\pi G\left({2\rho\over c^2}\right)\;.\label{eq:Poisson1}
\end{equation}
The factor 2 multiplying the density is directly due to the momentum within the source, 
which cannot be ignored when the gravitating particles cannot come to rest. When the 
source of gravity is radiation, Newton's `gravitational mass' $M_g$ producing 
the potential in Equation~(\ref{eq:Phi}) must therefore be twice the value one
would naively have calculated from the energy density alone. (But beware that this
is not the factor 2 associated with the deflection of starlight by the Sun. A correction
such as this, in cases where the source of gravity is not completely inertial, is
typically absorbed into the empirically determined value of $M_g$.)

Once the potential $\Phi$ and the Newtonian gravitational mass $M_g$ have been
properly identified in Equation~(\ref{eq:Phi}), we may then proceed with
Equation~(\ref{eq:geo}) to derive the correct equations of motion for a photon 
in the vicinity of a spherically symmetric object, where the appropriate 
spacetime is described by the Schwarzschild metric:
\begin{equation}
\hskip-0.7in ds^2=\left(1-{r_{\rm S}\over r}\right)c^2\,(dt)^2-
\left(1-{r_S\over r}\right)^{-1}(dr)^2-
r^2\left[(d\theta)^2+\sin^2\theta\,(d\phi)^2\right]\;.\label{eq:Schwarzschild}
\end{equation}
In this expression,
\begin{equation}
r_{\rm S}\equiv {2GM_g\over c^2}\label{eq:Sch}
\end{equation}
is the Schwarzschild radius for an object with gravitational mass $M_g$. Our goal
is to uncover the radial acceleration experienced by the photon at a radius
$r\gg r_{\rm S}$, i.e., in the weak-field limit and, to facilitate the calculation, 
we shall assume that the photon's velocity is perpendicular to ${\bf r}=r\hat{r}$ at 
that instant. 

The non-zero Christoffel symbols for this metric are simply
\begin{eqnarray}
{\Gamma^r}_{tt}&=&{1\over 2}{r_{\rm S}\over r^2}\left(1-{r_{\rm S}\over r}\right)\nonumber \\
{\Gamma^r}_{rr}&=&-{1\over 2}{r_{\rm S}\over r^2}\left(1-{r_{\rm S}\over r}\right)^{-1}\nonumber \\
{\Gamma^t}_{tr}&=&{1\over 2}{r_{\rm S}\over r^2}\left(1-{r_{\rm S}\over r}\right)^{-1}=
{\Gamma^t}_{rt}\nonumber \\
{\Gamma^\theta}_{r\theta}&=&{1\over r}={\Gamma^\theta}_{\theta r}\nonumber \\
{\Gamma^r}_{\theta\theta}&=&-r\left(1-{r_{\rm S}\over r}\right)\nonumber \\
{\Gamma^\phi}_{r\phi}&=&{1\over r}={\Gamma^\phi}_{\phi r}\nonumber \\
{\Gamma^r}_{\phi\phi}&=&-r\sin^2\theta\left(1-{r_{\rm S}\over r}\right)\nonumber \\
{\Gamma^\theta}_{\phi\phi}&=&-\sin\theta\cos\theta\nonumber \\
{\Gamma^\phi}_{\theta\phi}&=&\cot\theta={\Gamma^\phi}_{\phi\theta}\;.\label{eq:ChristSch}
\end{eqnarray}
Thus, letting overdot denote differentiation with respect to $\lambda$, we obtain
the following expression from the $\mu=0$ component of Equation~(\ref{eq:geo}):
\begin{equation}
\ddot{t}+\dot{t}\dot{r}{r_{\rm S}\over r^2}\left(1-{r_{\rm S}\over r}\right)^{-1}=0\;.\label{eq:one}
\end{equation}
Integrating this equation once gives
\begin{equation}
\dot{t}\left(1-{r_{\rm S}\over r}\right)=K_1\;,\label{eq:two}
\end{equation}
where $K_1$ is a constant of integration.

The $\mu=2$ component gives
\begin{equation}
\ddot{\theta}+{2\over r}\dot{r}\dot{\theta}-{\dot{\phi}}^2\sin\theta\cos\theta=0\;,\label{eq:three}
\end{equation}
while $\mu=3$ results in
\begin{equation}
\ddot{\phi}+{2\over r}\dot{r}\dot{\phi}+2\dot{\theta}\dot{\phi}\cot\theta=0\;.\label{eq:four}
\end{equation}
We align our coordinate system so that the photon's trajectory is restricted to the equatorial 
plane. Then, $\theta(\lambda)=\pi/2$ and $\dot{\theta}(\lambda)=0$, and Equation~(\ref{eq:three}) 
becomes irrelevant. Equation~(\ref{eq:four}) reduces to
\begin{equation}
\ddot{\phi}+{2\over r}\dot{r}\dot{\phi}=0\;,\label{eq:five}
\end{equation}
whose solution is simply
\begin{equation}
r^2\dot{\phi}=K_2\;,\label{eq:six}
\end{equation}
where $K_2$ is a second constant of integration.

Finally, each tangent vector along a null geodesic is light-like, so 
\begin{equation}
g_{\alpha\beta}{dx^\alpha\over d\lambda}{dx^\beta\over d\lambda}=0\;,\label{eq:seven}
\end{equation}
which provides us with the last equation we need:
\begin{equation}
\left(1-{r_{\rm S}\over r}\right)c^2{\dot{t}}^2-\left(1-{r_{\rm S}\over r}\right)^{-1}{\dot{r}}^2
-r^2{\dot{\phi}}^2=0\;.\label{eq:eight}
\end{equation}
Now using Equations~(\ref{eq:two}) and (\ref{eq:six}) with a change in variable
to $u\equiv 1/r$, together with 
\begin{equation}
{dr\over d\lambda}={dr\over d\phi}{d\phi\over d\lambda}\label{eq:nine}
\end{equation}
(which is valid because the motion has no $\theta$ dependence), 
we can modify Equation~(\ref{eq:eight}) to read as follows:
\begin{equation}
c^2K_1^2-K_2^2\left({du\over d\phi}\right)^2-K_2^2u^2\left(1-r_{\rm S}u\right)=0\;.\label{eq:ten}
\end{equation}
Differentiating this equation once more with respect to $\phi$ gives us the equation of
motion for a photon,
\begin{equation}
{d^2u\over d\phi^2}+u={3\over 2}r_{\rm S}u^2\;,\label{eq:eleven}
\end{equation}
which may be thought of as a relativistic version of the classical Binet orbit equation 
\cite{DEliseo:2007}, best known for its use in calculating the deflection angle of 
starlight grazing the surface of the Sun on its way toward Earth.

In the absence of gravity ($r_{\rm S}\rightarrow 0$), the solution to this equation
would be a straight line perpendicular to {\bf r}, given as
\begin{equation}
u={1\over r_0}\cos\phi\;,\label{eq:twelve}
\end{equation}
where $r_0$ is the radius of closest approach to the origin, corresponding to
the value of $r$ at $\phi=0$. If we now make a weak-field approximation, consistent
with $r\gg r_S$, the corresponding null trajectory will be a perturbation of this 
straight line, allowing us to write
\begin{equation}
{d^2u\over d\phi^2}+u={3\over 2}{r_{\rm S}\over r_0^2}\cos^2\phi\;,\label{eq:thirteen}
\end{equation}
whose solution is 
\begin{equation}
u={r_{\rm S}\over r_0^2}+{1\over r_0}\cos\phi-{1\over 2}{r_{\rm S}\over 
r_0^2}\cos^2\phi\;.\label{eq:fourteen}
\end{equation}

Our interest here is not so much how the null geodesic is modified due to the
accumulated effect of gravity over the ray's transit to Earth but, rather, the
instantaneous acceleration experienced by a photon in the vicinity of $\phi=0$. 
The acceleration is obtained by differentiating this expression 
twice with respect to the observer's time, $t$, starting with 
\begin{equation}
-{1\over r^2}{dr\over dt}=-{\sin\phi\over r_0}{d\phi\over dt}+{r_{\rm S}\over
r_0^2}\cos\phi \sin\phi {d\phi\over dt}\;.\label{eq:fifteen}
\end{equation}
We also have
\begin{equation}
r{d\phi\over dt}=c\cos\phi\;,\label{eq:sixteen}
\end{equation}
and so
\begin{equation}
{dr\over dt}=c\cos\phi-{cr_{\rm S}\over r}\sin\phi\;,\label{eq:seventeen}
\end{equation}
using also the relation in Equation~(\ref{eq:twelve}), since gravity perturbs
this trajectory only slightly in the weak-field limit.

A second differentiation results in the expression
\begin{equation}
{d^2r\over dt^2}={c^2\over r}\cos^2\phi+{cr_{\rm S}\over r^2}\sin\phi{dr\over dt}
-{c^2r_{\rm S}\over r^2}\cos^2\phi\;.\label{eq:eighteen}
\end{equation}
In this equation, however, the first term is simply the effect of seeing the
straight line trajectory (Equation~\ref{eq:twelve}) in spherical coordinates.
The radial acceleration due to gravity in Equation~(\ref{eq:eighteen}) is the 
rest of the righthand side. Thus, subtracting the first (geometric) term,
substituting for $dr/dt$ from Equation~(\ref{eq:seventeen}), and noting that
$r_{\rm S}\ll r$ in the weak-field limit, we arrive at the Newtonian acceleration
for a photon
\begin{equation}
{d^2r\over dt^2}=-{2GM_g\over r^2}\left(1-2\sin^2\phi\right)\;.\label{eq:nineteen}
\end{equation}

The dependence of this expression on the angle $\phi$ is simply due to the
fact that the photon experiences zero acceleration in its longitudinal direction,
so Equation~(\ref{eq:nineteen}) gives solely the component of acceleration in
the radial direction. Be aware, however, that the approximations we have made
in reaching this result are valid only near $\phi=0$, so this expression is not
valid when $\sin\phi\rightarrow 1$. On the other hand, given that $\phi\rightarrow 0$ 
when $r\rightarrow r_0$, this simply reduces to our final result,
\begin{equation}
{d^2r\over dt^2}=-{2GM_g\over r^2}\;,\label{eq:Newtonnull}
\end{equation}
which is the maximal radial acceleration experienced by a photon transverse to 
its direction of motion.

Equation~(\ref{eq:Newtonnull}) is the weak-field ($r\gg r_{\rm S}$), radial acceleration 
experienced by a photon moving perpendicular to $\hat{r}$ in the vicinity of a gravitational 
mass $M_g$. It represents the Newtonian, static limit of General Relativity for particles
we believe to have zero rest mass, the analog of Equation~(\ref{eq:Fg}).  Aside from their evident 
similarity, the other feature that stands out is the additional factor 2 emerging in 
the latter, which owes its appearance to the same physics responsible for the famous 
factor 2 in the calculation of the deflection angle of starlight grazing the surface 
of the Sun. This factor 2 appears as long as we use the same value of
the gravitational constant, $G$, in both Equations~(\ref{eq:Fg}) and (\ref{eq:Newtonnull}).
It is not due to a doubling of the source in Poisson's Equation~(\ref{eq:Poisson1}),
which arises from the contribution of momentum to the active gravitational mass, and
would be absorbed into the overall `measured' value of $M_g$, independently of $G$. 

It is instead due to the different geometries of particles with and without inertia
along their longitudinal direction of motion. Simply put, particles with inertia 
satisfy Equation~(\ref{eq:Fg}) in the low-velocity limit because, for them, the only 
metric coefficient in the Schwarzschild spacetime (Eq.~\ref{eq:Schwarzschild}) that 
matters is $g_{tt}$, i.e., the time dilation. As noted earlier in this 
section, Einstein himself used solely the effects of $g_{tt}$ to estimate the bending 
of light passing near the Sun, even though a lightwave was thought to be massless. 
These particles are moving too slowly compared to the speed of light for the distance 
covered during their acceleration to contribute significantly to $ds^2$. For particles 
propagating at lightspeed, however, the impact of $g_{rr}$ cannot be ignored compared 
to $g_{tt}$. In essence, the effects of spacetime curvature are doubled for photons 
compared to particles with established non-zero inertia.  The inclusion 
of spatial variations resulting from $g_{rr}$, once the full theory of general 
relativity was available to him, is the reason Einstein's recalculation of the 
deflection angle of starlight passing near the Sun doubled the effect he had 
anticipated in 1911.

\section{Mass and Rest-mass Energy}\label{mgmi}
The inference we draw from Equations~(\ref{eq:gPhi}), (\ref{eq:Poisson}), 
(\ref{eq:Poisson1}) and (\ref{eq:Newtonnull}) is that all particles experience
a Newtonian-like gravitational attraction to each other in the non-relativistic 
limit, whose coupling strength---according to general relativity---is the total
energy in each of the gravitating objects. The Newtonian approach of assigning 
them `masses' appears to be a way of representing these energies in terms of an 
inertial or gravitational context, a distinction that no doubt arises from the 
nature of rest-mass energy, as we shall further develop in this section. 

In our previous, detailed examination of the origin of rest-mass energy
\cite{Melia:2021a}, we showed that, of the four known forces, only gravity has 
all of the attributes required to satisfy a `Principle of Equivalence'. Thus, 
the energy we commonly assign to a particle's rest mass is almost certainly 
gravitational in nature. Indeed, in the context of modern cosmology, the 
binding energy of a particle with gravitational mass $m_g$ in causal
contact with that portion of the Universe within our gravitational horizon,
$R_{\rm h}\equiv c/H$, where $H$ is the Hubble constant, is exactly $m_gc^2$.  
If inertial mass is viewed as a surrogate for $m_g$, we find in this result 
a natural explanation for the origin of rest-mass energy, though it still 
leaves open the question of whether $m_i$ and $m_g$ are truly different 
characteristics of the same object (or particle). 

We should stress at this point that the proposal being discussed
here, and introduced in ref.~\cite{Melia:2021a}, is unique in a cosmological
setting for the simple reason that it directly addresses the question of
where rest-mass energy comes from, not simply mass. For example, other
definitions of mass in cosmology, such a the Komar mass \cite{Komar:1963}, 
or the Tolman mass \cite{Tolman:1930}, are statements concerning how much 
`mass' is required to account for the (static) spacetime curvature in a 
closed volume, but none of these alternative approaches explains why the 
energy associated with $m_g$ must be $m_gc^2$. In our proposal, on the other 
hand, any particle with gravitational mass $m_g$ has a binding energy 
$m_gc^2$ due to its gravitational coupling to the energy contained within 
$R_{\rm h}$. 

What we learn from Equation~(\ref{eq:Newtonnull}) is that
particles we believe to have zero rest mass nevertheless also experience a 
gravitational force that accelerates them at a {\it finite} rate, albeit 
solely in a direction perpendicular to their velocity (with fixed magnitude 
$c$). Attempting to naively interpret this result in the context 
of Newton's original formulation of his law of universal gravitation 
(Eq.~\ref{eq:Fg}), however, we would instead be compelled to assign 
them a non-zero inertial `mass', for otherwise their acceleration would 
be infinite. Of course, this simple-minded approach does not comport very 
well with our conventional view that photons should be `massless.' It 
appears, therefore, that our current definition of mass, inertial or 
otherwise, may be inaccurate, perhaps even defective. 

Let us reconsider the Newtonian gravitational attraction between two
particles. The clue we glean from general relativity, e.g., via Equation~(\ref{eq:trace}), 
is that the coupling constant for this interaction is
the particle's total energy, $E$. And since the Newtonian gravitational force
is symmetric between them, we write the force on particle 1 due to 2 as
\begin{equation}
{\bf F}_{g1}= -{Gm_{g1}m_{g2}\over r^2}\hat{r}\;,\label{eq:Fg1}
\end{equation}
where {\bf r} points from 1 to 2.  Crucially, we now define
\begin{equation}
(m_gc^2)^2\equiv E^2=(m_ic^2)^2+(qc)^2\;,\label{eq:mgenergy}
\end{equation}
where $q$ here is the momentum of particle $m_i$ (distinguishing it from the
symbol $p$ we used earlier to denote the pressure). 

According to our earlier study on the origin of rest-mass energy,
\cite{Melia:2021a}, the binding energy of particle $m_g$ to that portion of
the Universe within $R_{\rm h}\equiv c/H$, which also coincides with our
apparent horizon \cite{Melia:2018b} and the Hubble radius, equals its escape 
energy $E_{\rm esc}=qc$ attained when its proper distance, $R$, approaches 
$R_{\rm h}$. Since $q\rightarrow m_ic$ in this limit, we infer that 
$E_{\rm esc}=m_ic^2$. In contrast, a photon always has escape speed, 
even at $R_{\rm h}$, and is therefore unbound. For such particles, the 
total energy at any $R$ is simply $E=qc$. 

Consequently, Equation~(\ref{eq:mgenergy}) may also be written 
\begin{equation}
(m_gc^2)^2\equiv E_{\rm esc}^2+(qc)^2\;,\label{eq:mgenergy1}
\end{equation}
though $E_{\rm esc}\rightarrow 0$ for photons. With the definition in 
Equations~(\ref{eq:mgenergy}) and (\ref{eq:mgenergy1}), photons are therefore
assigned a `gravitational mass'
\begin{equation}
m_g^\gamma\equiv{q\over c}\;,\label{eq:mgphoton}
\end{equation}
while matter has the corresponding value 
\begin{equation}
m_g^{\rm m}=\left[(m_i^{\rm m})^2+\left({q\over c}\right)^2\right]^{1/2}\;.\label{eq:mgenergy2}
\end{equation}
Newton's law of universal gravitation is valid only in the low-velocity limit, however,
for which $qc\ll m_i^{\rm m}c^2$, and therefore
\begin{equation}
m_g^{\rm m}\rightarrow m_i^{\rm m}\;,\label{eq:mgenergy3}
\end{equation}
fully consistent with the Principle of Equivalence.

But what we have learned from Equation~(\ref{eq:Newtonnull}) is that a photon must 
also experience the force in Equation~(\ref{eq:Fg1}) in the Newtonian limit, albeit
restricted to a direction perpendicular to its velocity.  Newton's equation of motion 
for such a particle should therefore be written 
\begin{equation}
m_i^\gamma{d^2{\bf r}\over dt^2}=-{2Gm_g^\gamma M_g\over r^2}\hat{r}\;,\label{eq:Newtphoton}
\end{equation}
implying that 
\begin{equation}
m_i^\gamma=m_g^\gamma\label{eq:Newtmi}
\end{equation}
for the photon as well, to be consistent with Equation~(\ref{eq:Newtonnull}).
Not surprisingly, this is what we should have expected from the Principle of
Equivalence applied to {\bf all} particles, not just matter. {\sl Note, however, 
that unlike the inertial mass of matter, $m_i^\gamma$ does not carry 
a `rest energy' because the photon is always unbound.}  As such, $m_i^\gamma$ 
does not appear in the photon's total energy budget analogous to 
Equation~(\ref{eq:mgenergy}).

\section{Conclusion}\label{conclusion}
An important caveat for this work is that much of our discussion hinges on 
the viability of general relativity as the correct description of nature. In 
particular, all of the results we have discussed stem directly or indirectly 
from the fact that the source of gravity in Einstein's theory is the total 
energy of the system. No experimental test has ever provided evidence against 
this feature. 

A consideration of Einstein's theory in the weak-field, static 
limit then yields a workable interpretation of Newton's more empirical law of
universal gravitation. We have shown in this paper that a direct comparison
of these two approaches provides us with a possible explanation for the 
physical origin of gravitational and inertial mass, and perhaps also
for the origin of rest-mass energy within the framework of one of the 
most famous solutions to Einstein's equations, i.e., the 
Friedmann-Lema\^itre-Robertson-Walker (FLRW) metric \cite{Melia:2020}. 

This work has also highlighted a tantalizing result that we should have 
anticipated all along. We have had an observational confirmation for 
several decades that astronomical sources of gravity bend null geodesics,
fully confirming an important prediction of Einstein's theory. But the implied
acceleration producing the deflection is finite, and when this is viewed in the
weak-field, static limit (i.e., within a Newtonian framework), we cannot
avoid the conclusion that light must also have a non-zero `inertia' 
associated with the acceleration perpendicular to its velocity. 

We should mention in passing that this notion of (at least
some) particles possessing a velocity-dependent (which in our case
translates into a direction-dependent) inertia echoes the work of Lorentz
and subsequent workers initiated in 1899 \cite{Lorentz:1899}. The basis
of his argument was the application of the nonrelativistic formula
${\bf p}=m{\bf v}$ in the relativistic domain, which we now know is
incorrect. Nevertheless, his influence at the turn of the century, prior
to Einstein's introduction of special relativity, was significant enough
to attract attention from the physics community to the broader question
of the meaning of mass. In fact, his 1904 paper `Electromagnetic Phenomena in a 
System Moving With Any Velocity Less Than That of Light' \cite{Lorentz:1904}
introduced the ``longitudinal'' and ``transverse'' electromagnetic masses of 
the electron. This view was eventually supplanted by Einstein's moving 
observer's inertia, $\gamma m$, however, so the notion of a direction-dependent
inertia eventually subsided, though some of his concepts are still being
considered today (see, e.g., ref.~\cite{Maurya:2016}).

In this paper, we have demonstrated that a direction-dependent inertia may
not be out of the question after all, at least for photons. Without question,
these particles have no inertia in the longitudinal direction, but they 
evidently do resist acceleration in the transverse direction. This inertia is 
apparently proportional (or even equal) to the gravitational mass inferred from 
the photon's energy but, at this stage, we have no idea what produces it. 
The speed of light is always $c$, however, so photons are always unbound within 
our gravitational horizon in the context of FLRW. This `transverse' inertia thus 
carries no energy, and is therefore absent from the expression yielding the 
photon's total energy budget. 

In summary, then, we have argued that gravitational mass, $m_g$, 
is---in all cases----a surrogate for the particle's total energy which, in the context 
of Einstein's theory, is the actual source of spacetime curvature. Inertia no longer
appears to be an intrinsic property but, rather, is an emergent feature due to 
several different mechanisms. For particles exhibiting a resistance to acceleration 
in their longitudinal direction of motion, this inertia appears to be proportional
to $m_g$. This conclusion is supported by the interpretation that rest-mass energy
is a binding energy within our gravitational (or Hubble) horizon in the context of 
the FLRW cosmic spacetime. A consideration of particle motion near this horizon
shows that the particle's energy approaches $E_{\rm esc}=qc$, with a momentum
$q=m_ic$, and Newton's law of universal gravitation in the low-velocity limit
then shows that $m_g\rightarrow m_i$ (where the equality ensues with an appropriate
choice of $G$). Photons exhibit an inertia transverse to their motion, and the
implied inertial mass equals their gravitational mass, consistent with the Principle
of Equivalence. But this inertia has no impact on their longitudinal motion, so
it does not affect their energy budget, which is entirely kinetic.

We may have generated more questions than answers with this discussion, but 
hopefully in a manner that encourages further attempts at uncovering the true,
physical relationship between a particle's inertial and gravitational mass(es). 

\section*{Data availability statement}
No new data were created or analyzed in this study.


\section*{References}

\bibliographystyle{iopart-num} 
\bibliography{ms.bib}

\end{document}